\documentclass[useAMS,usenatbib]{mn2e}

\usepackage{amsmath}
\usepackage{amssymb}
\usepackage{mhchem}
\usepackage{color}
\usepackage{graphicx}
\usepackage{natbib}
\usepackage{pdflscape}
\usepackage{hyperref}
\usepackage{caption}

\title{O, Na, Ba and Eu abundance patterns in open clusters}
\author[B. T. MacLean, G. M. De Silva and J. Lattanzio]{B. T. MacLean$^{1}$\thanks{E-mail:
btmaclean@gmail.com; gayandhi.desilva@aao.gov.au; john.lattanzio@monash.edu}, G. M. De Silva$^{2}$\footnotemark[1] and J.
Lattanzio$^{1}$\footnotemark[1]\\
$^{1}$Monash Centre for Astrophysics (MoCA), Monash University, Victoria 3800, Australia\\
$^{2}$Australian Astronomical Observatory, 105 Delhi Rd, North Ryde, NSW 2113, Australia}
\begin{document}

\date{Accepted TBC. Received TBC; in original form TBC}

\pagerange{\pageref{firstpage}--\pageref{lastpage}} \pubyear{2014}

\maketitle

\label{firstpage}

\begin{abstract}
Open clusters are historically regarded as single-aged stellar populations representative of star formation within the Galactic disk. Recent literature has questioned this view, based on discrepant Na abundances relative to the field, and concerns about the longevity of bound clusters contributing to a selection bias: perhaps long-lived open clusters are chemically different to the star formation events that contributed to the Galactic disk. We explore a large sample of high resolution Na, O, Ba \& Eu abundances from the literature, homogenized as much as reasonable including accounting for NLTE effects, variations in analysis and choice of spectral  lines. Compared to a template globular cluster and representative field stars, we find no significant abundance trends, confirming that the process producing the Na-O anti-correlation in globular clusters is not present in open clusters. Furthermore, previously reported Na-enhancement of open clusters is found to be an artefact of NLTE effects, with the open clusters matching a subset of chemically tagged field stars.

\end{abstract}



\section{Introduction}

Many spectroscopic studies support that Galactic open clusters (OCs), both young and old, host stars with chemically similar properties, having undergone a single burst of star formation from a highly homogenized progenitor cloud \citep[e.g.][]{desilva2006,randich2006study,desilva2007,pancinocarrera1,magrini2014}. Dissipation of star clusters is thought to be a major contributor to building up the Galactic thin disk, and attempts have been made to reconstruct some of these dispersed clusters from the local field using the method of chemical tagging \citep{chemtagdata}.

Globular clusters (GCs) on the other hand display strong evidence of undergoing more than one stellar birth event, with self-pollution by an earlier generation of stars \citep{cottrell1981}. Evidence for this theory exists in the photometric and spectroscopic observations and perhaps most notably in the Na-O anti-correlation seen across both evolved and unevolved members \cite[and references therein]{gratton2004,gratton2012}. There is evidence for other light elemental abundance variations in GCs, however GCs are (mostly) observed to be homogeneous in iron and the heavier elements \citep{suntzeff1993,gratton2004,carretta2009}.

It has long been hypothesized that the Galactic halo is partly composed of stars that were stripped from GCs during the early history of the Galaxy. However it is also known that very few stars in the halo show the abundance patterns seen in present-day GCs. Recent studies show that only about 3 $\pm$ 2 per cent of halo stars show either the CN variations seen in GCs \citep[e.g.][]{carretta2010,martell2010,martell2011} or the Na-O anti-correlations that are ubiquitous in GC stars \citep{ramirez2012}. This is actually consistent with the current preferred scenario for the formation history of the Galaxy and the multiple population scenario in GCs. In this model the proto-GCs are much more massive than their present-day manifestations, by factors of 10 to 20 or so. It is preferentially the first generation of stars in the GCs (but also some second generation), that are lost from the proto-GC and go on to form the halo of the Galaxy.

It can be argued that GCs universally show the Na-O anti-correlation (a few possible exceptions are Ter 7 \citep{ter7a,ter7b}, Pal 12 \citep{pal12} and Rup 106 \citep{rup106}), with \cite{carretta2010} suggesting that the definition of a bona fide GC is one that shows this pattern. Several studies have searched for signatures of the Na-O anti-correlation in OCs, so far with negative results. But high levels of Na in OCs compared to the field have been noted by many authors \citep[e.g.][and references therein]{desilva2009oc,pancinocarrera2}, but not in every analysis \citep[e.g.][]{pancinocarrera1}, and with \cite{smiljanic2012} suggesting that these results may have arisen from NLTE and evolutionary mixing effects.

This questions if presently bound open clusters are valid examples of typical star formation events that contributed to the Galactic disk. Dynamically, the older bound clusters, which have deeper potentials, may be anomalous objects that faced a different formation and chemical evolution, where embedded clusters subject to high infant mortality rates are the major contributors to the building of the Galactic disk \citep{lada2003}. 

In this paper we explore the abundances of Na, O, Ba and Eu in OCs by homogenizing literature studies as much as reasonable \citep[a similar study on Fe abundances in OCs in the literature was done by][]{heiter2014}. These elements were chosen as they represent the extremes in observed abundance scatter; Na and O showing the largest scatter and the neutron capture elements Ba (s-process) and Eu (r-process) being the most homogeneous. We compare the resulting dataset against a typical GC anti-correlation as well as against the chemically tagged coeval clusters found within the disk abundances by \cite{bensby2014study}. 


\section{Method}


\subsection{Data collection}

We sourced high resolution spectroscopic abundances of Na and O  as well as Ba and Eu in Galactic open clusters from the literature, in order to gather the largest sample size possible. If a study did not measure either pair of these elements, it was not included in our sample. To minimise systematic effects and ensure a high level of abundance accuracy, we limited our data set to studies based on a minimum spectral resolution of R = 20,000. Due to the fact that the anti-correlations in globular clusters and chemical homogeneity in open clusters are observed as star-to-star abundance variations, we originally limited our open cluster sample to clusters where a total of five or more individual stellar abundances were available. However due to the scarcity of studies that measured both Ba and Eu, the latter restriction was not applied to these two elements. No selection was made on stellar type because the GC anti-correlation is observed in unevolved dwarf stars as well as in giant stars \citep{carretta2004, gratton2001}, but their effects on abundance results were taken into account when exploring systematic uncertainties (see Section \ref{homo}). 

The final total sample consists of 228 stars (207 with Na \& O abundances, and 40 with Ba \& Eu abundances) in 19 open clusters, with the reference list of included data given in Table \ref{tab:refs}. As a template for the well known GC anti-correlation, Na and O abundances for NGC 2808 were adopted from  \cite{carretta2006a}. For a comparison to the Galactic field, we used the latest disk sample by \cite{bensby2014study} for Na \& O abundances, and \cite{bensby2004study} for Ba \& Eu abundances.

\begin{table}
\caption{The reference list of open cluster survey data used in the sample.}
\label{tab:refs}
\begin{tabular}{|l|l|}
\hline
Population & Reference  \\ \hline
Be 39 & \cite{bragaglia2012study} \\
Collinder 261 & \cite{friel2003study} \\
 & \cite{carretta2005study} \\
Hyades & \cite{varenne1999study} \\
 & \cite{schuler2006study} \\
 & \cite{schuler2009study} \\
M 67 & \cite{tautvaisiene2000study} \\
 & \cite{randich2006study} \\
 & \cite{pace2008study} \\
Melotte 111 & \cite{gebran2008study} \\
NGC 752 & \cite{reddy2012study} \\
NGC 1817 & \cite{reddy2012study} \\
NGC 2360 & \cite{reddy2012study} \\
NGC 2506 & \cite{mikolaitis2011bstudy} \\
 & \cite{reddy2012study} \\
NGC 3114 & \cite{santrich2013study} \\
NGC 6134 & \cite{mikolaitis2010study} \\
NGC 6253 & \cite{carretta2007study} \\
NGC 6475 & \cite{villanova2009study} \\
NGC 6791 & \cite{carretta2007study} \\
 & \cite{geisler2012} \\
NGC 7789 & \cite{tautvaisiene2005study} \\
IC 4651 & \cite{mikolaitis2011study} \\
Pleiades & \cite{gebran2008bstudy} \\
Praesepe & \cite{boesgaard2013study} \\
Trumpler 20 & \cite{carraro2014study} \\
NGC2808 & \cite{carretta2006a} \\
Local Field & \cite{bensby2004study} \\
 & \cite{bensby2014study} \\ \hline
\end{tabular}
\end{table}


\subsection{Homogenization of data}\label{homo}

Each study carries along with it its own method for abundance analysis, systematic errors and reference scale. This section describes the various differences in analysis and the steps taken in homogenising the collated data. The final adjusted data set covering all elements per star and literature source is presented in Table \ref{tab:sample}.

As part of the data homogenization process, the first step was to set a standard solar reference scale where possible. The published abundances were normalized to the solar values given by \cite{asplund2009}, where the reference abundance is log $\epsilon_{Fe}$ = 7.50, log $\epsilon_{O}$ = 8.69, log $\epsilon_{Na}$ = 6.24, log $\epsilon_{Ba}$ = 2.18 and log $\epsilon_{Eu}$ = 0.52. This normalization was applied to all stars in our sample, except for stars from studies that carried out a differential analysis with respect to the Sun. Such were the OC studies by \citet[Hydes]{schuler2009study}; \citet[M67]{tautvaisiene2000study}; \citet[NGC7789]{tautvaisiene2005study}; \citet[NGC6134]{mikolaitis2010study}; \citet[IC4651]{mikolaitis2011study}; and \citet[NGC2506]{mikolaitis2011bstudy}.

\subsubsection{O abundances}

In the derivation of O abundances, most studies consistently used spectrum synthesis on one or both of the [O I] forbidden lines at 6300.3\AA\ and 6363.8\AA. \cite{tautvaisiene2000study,randich2006study}; and \cite{pace2008study} carried out an equivalent width (EW) analysis of the line at 6300.3\AA\ (in all cases on M67 stars), resulting in typical measurement uncertainties 0.05 dex larger than those studies that used spectrum synthesis, presumably due to unaccounted effects of line blending.

\citet[Mel111]{gebran2008study}; \citet[Pleiades]{gebran2008bstudy}; and \citet[Hyades]{varenne1999study} applied spectral synthesis on the O triplet lines at 6155\AA, 6156\AA\ and 6158\AA, which are affected by departures from LTE \citep{takeda2006} but were not accounted for. These were also the only studies whose sample include A-type dwarf stars, which show considerable scatter in [O/Fe], with an average standard deviation of 0.24 dex compared to 0.10 dex in the rest of the sample.

The field sample by  \cite{bensby2014study} was subject to a differential solar analysis, where the oxygen abundances were based on the triplet at 7774\AA. Their measured solar oxygen values were approximately 0.15 dex larger than that of \cite{asplund2009} adopted in this study. Given the differential nature of the field sample, we adopt the \cite{bensby2014study} abundances as published.

\subsubsection{Na abundances}

For the analysis of Na, most studies used EWs to derive abundances. Exceptions to this were \cite{gebran2008study,gebran2008bstudy,varenne1999study}; and \citet[NGC6253]{carretta2007study}, who used full spectrum synthesis to derive all abundances, mostly based on \cite{takeda1995}. It is well known that Na lines are subject to non-local-thermodynamic-equilibrium (NLTE) effects \citep{takeda2003,lind2011,gratton1999}, which can cause large variations in Na abundance compared to the results if LTE is assumed. 

While the NGC 2808 stars and disk field samples included NLTE corrections, only 60 per cent of OC studies had accounted for NLTE effects. For Na abundances which were based on LTE assumptions, these were adjusted to correct for NLTE effects as described in \cite{lind2011} by using the web-based INSPECT interface \footnote{http://inspect-stars.net}. The system takes as input the stellar parameters (metallicity, effective temperature, surface gravity and microturbulence) together with either the EW or Na-LTE abundance for a particular line. \cite{gebran2008study}  and \cite{schuler2009study} were the only studies to provide EWs per line. For the remaining LTE-based OC sample, the correction for each star was derived using the average Na abundance as input for all lines because the Na abundances for each individual line were not provided by the literature source.  Assuming the same Na abundance for each line is strictly incorrect, however the standard deviation of the correction across several lines is noted to be less than 0.06 in all cases.

For the studies that accounted for NLTE effects, we explored the impact of their applied corrections as different prescriptions could lead to different corrections. For Be 39, \cite{bragaglia2012study} based their corrections on the prescription by \cite{lind2011}. For NGC 6475, \cite{villanova2009study} used the corrections from \cite{gratton1999} to correct for Na NLTE effects. We use their stated Na LTE abundances to estimate the corrections based on \cite{lind2011} via the INSPECT web-based interface. We find that for the cooler stars the corrections applied by \cite{villanova2009study} are comparable to those estimated via the INSPECT queries to within 0.01 dex. Corrections for the warmer stars are beyond the grid provided by \cite{lind2011}. 

The studies by \cite{carretta2005study} on Collinder 261, \cite{mikolaitis2011study} on IC 4651, \cite{mikolaitis2010study} on NGC 6134 and \cite{santrich2013study} on NGC 3114 also followed  \cite{gratton1999} to estimate the NLTE correction. \cite{lind2011} notes that the increasing trend towards positive correction values for very low surface gravity stars seen in  \cite{gratton1999}, is not seen in their calculations. As these studies do not provide the Na LTE abundances or the line EWs, it is not possible for us to assess the differences in the applied correction. Judging by the case of NGC 6475 by  \cite{villanova2009study} (discussed above) the differences should be minor. In these studies an average of -0.2 dex is applied to Na abundances, compared to our average adjustment of -0.11 dex based on \cite{lind2011}.

Very little information exists on the NLTE effect on Na lines in hot A-type stars, therefore the Na values of these stars in \cite{varenne1999study,gebran2008study}; and \cite{gebran2008bstudy} were uncorrected.

\subsubsection{Ba and Eu abundances}
All studies included in our sample used spectrum synthesis for Ba and Eu abundances measurements, using all or a subset of the barium lines at 5853\AA, 6141\AA, 6496\AA\ and the europium line at 6645\AA, adopting similar atomic data. Many of these lines are affected by hyper-fine structure and isotopic shifts, which were included in the original studies. However subtle difference in methodology are likely to exist, which are difficult to pin down and are represented in the rms error about the cluster mean given in Table \ref{tab:sample}.

\begin{table*}
\centering
\caption{Open cluster sample, with star names as given in each paper and references for each star. The last row of every cluster is the average abundances with rms error. Abundances are the normalized values, corrected for NLTE effects. Column 5 indicates the NLTE correction on Na abundances, calculated as per \protect\cite{lind2011} and described in Section \ref{homo}. Full table available online.}
\label{tab:sample}
\begin{tabular}{|c|c|c|c|c|c|c|c|c|c|c|}
\hline
Population & Star & [O/Fe] & [Na/Fe] & NLTE Correction & [Ba/Fe] & [Eu/Fe] & Reference \\ \hline
Be 39 & 1256 & 0.22 $\pm$ 0.10 & -0.11 $\pm$ 0.00 & - & - & - & \cite{bragaglia2012study} \\ 
Be 39 & 1407 & 0.17 $\pm$ 0.10 & -0.02 $\pm$ 0.02 & - & - & - & \cite{bragaglia2012study} \\ 
Be 39 & 1657 & 0.28 $\pm$ 0.10 & -0.15 $\pm$ 0.03 & - & - & - & \cite{bragaglia2012study} \\ 
Be 39 & 2130 & 0.32 $\pm$ 0.10 & -0.09 $\pm$ 0.01 & - & - & - & \cite{bragaglia2012study} \\ 
Be 39 & 2144 & 0.20 $\pm$ 0.10 & -0.11 $\pm$ 0.01 & - & - & - & \cite{bragaglia2012study} \\ \hline
\end{tabular}
\label{}
\end{table*}


\section{Results and discussion}

\subsection{Cluster inhomogeneities}\label{inhomo}

In Figure~\ref{fig:naostars}, we plot the cluster average [Na/Fe] and [O/Fe] abundances (with standard deviations) from our sample of 14 open clusters against the results for the globular cluster NGC 2808. Only cluster average values are plotted because most clusters show homogeneity in [Na/Fe] and [O/Fe], representative of single stellar populations where the standard deviation of the data set is of a similar value (or smaller) than that expected to arise from typical measurement uncertainties (see Table \ref{tab:sample}).

Five clusters, however, appear to show larger star-to-star scatter than is predicted by uncertainties in measurements. These are Melotte 111, Hyades, NGC 6475, NGC 6791 and Pleiades, all which showed large abundance variations in O and Na; standard deviations for these clusters were on average 0.15 dex larger than their typical measurement errors. These apparent inhomogeneities, however, can all be shown to be artefacts arising from systematic errors in abundance analysis and stellar type. 

Of these apparently inhomogeneous clusters, O was calculated using NLTE affected lines for the stars in Melotte 111, Pleiades, and the Hyades \citep[from][]{varenne1999study} as discussed in Section \ref{homo}, whereas all other studies used the forbidden line of 6300.3\AA. These are also the only studies whose samples include A-type dwarf stars, which show considerable scatter, and for which Na abundances could not be corrected for NLTE effects.

In the case of NGC 6475 by  \cite{villanova2009study} and the Hyades abundances by \cite{schuler2009study}, it is a disparity between giant and dwarf stars that creates the apparent inhomogeneities; in both cases the dwarfs alone are homogeneous. This suggest that individual stellar evolutionary effects are the cause and the clusters when formed were homogeneous. 

Finally, NGC 6791 is noted by \cite{geisler2012} to show a trend in [Na/Fe] vs. [O/Fe]. Along with \cite{carraro2006} they question whether NGC 6791 is truly an OC, but a globular cluster or tidally disrupted and cannibalized dwarf galaxy. If it is an OC, it is most certainly a peculiar one; very old but metal-rich, and containing an unusually large population of binary systems and an extremely blue horizontal branch. We note however that new results by \cite{bragaglia2014} claim no evidence of a Na-O anti-correlation trend in this cluster.

\subsection{Na-O anti-correlation}

\begin{figure}
\centering
\includegraphics[width=9cm]{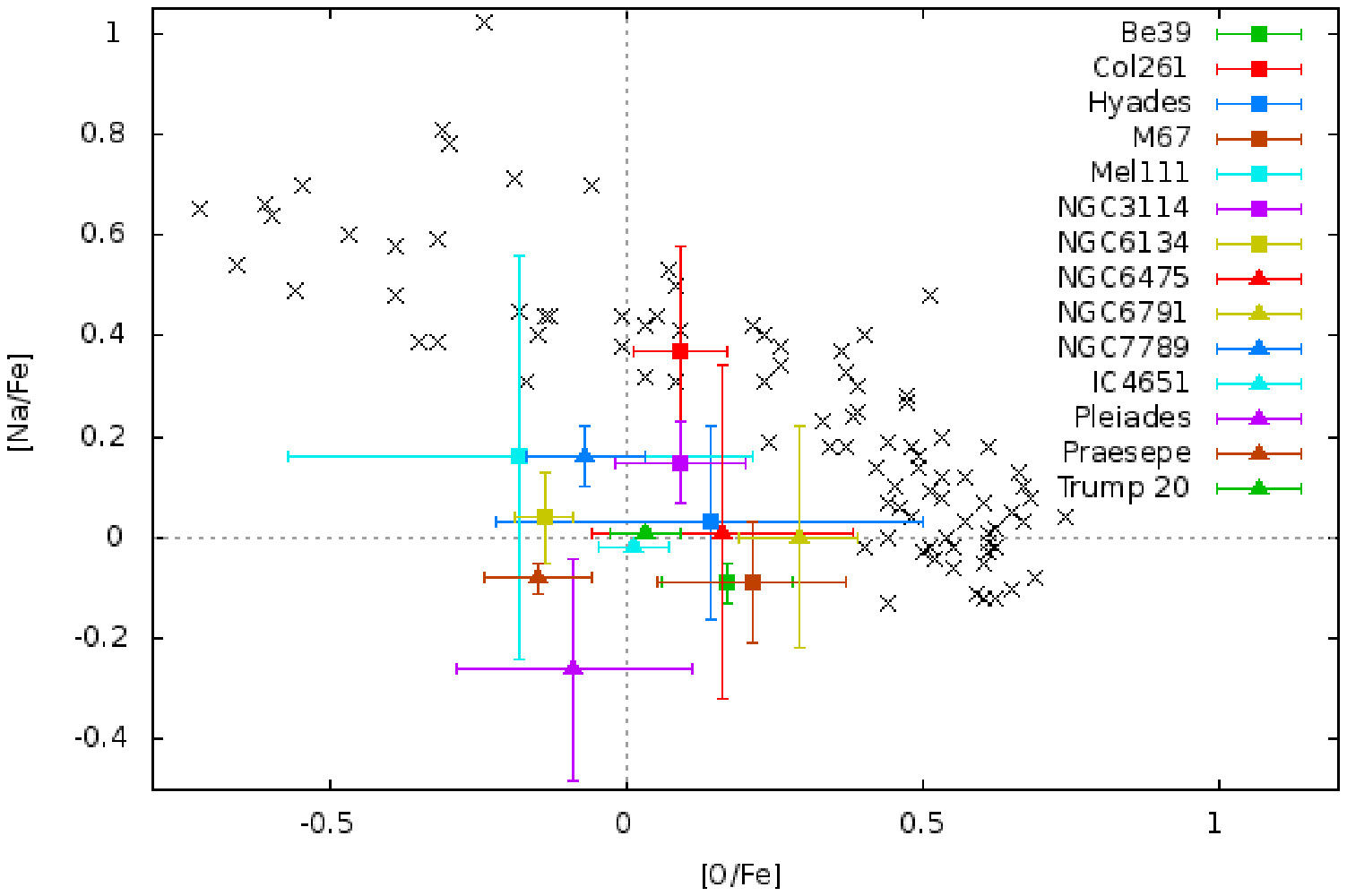}
\caption{Average normalized [Na/Fe] vs. [O/Fe] abundances with rms errors for open clusters (squares and triangles) and globular cluster NGC 2808 \protect\citep[black crosses, from][]{carretta2006a}. The large scatter of several clusters is discussed in Section \ref{inhomo}.}
\label{fig:naostars}
\end{figure}

As seen in Figure~\ref{fig:naostars}, the open clusters do not follow the clear anti-correlation seen in NGC 2808 for [Na/Fe] vs. [O/Fe]. The OC abundances do not reach either the extreme Na enhancements or the extreme O depletions seen in NGC 2808, and show values typical of disc stars. Thus we conclude that not only is there no Na-O anti-correlation from star-to-star within a given open cluster, but the cluster means also do not show such a correlation. In both cases the abundances are more typical of the disc stars.

Figure~\ref{fig:chemtag} shows the average normalized [Na/Fe] and [O/Fe] abundances of the homogenized open clusters with rms errors, along with representative field stars. In the calculation of the OC averages, those studies with large inconsistencies as discussed in Section \ref{inhomo}, were removed. These are all stars of Melotte 111 and Pleiades, the Hyades sample by \cite{varenne1999study}, the giant Hyades stars from \cite{schuler2006study} and \cite{schuler2009study}, and the giant stars in NGC 6475. \cite{chemtagdata} used the field sample by \cite{bensby2014study} to reconstruct dispersed clusters using the method of chemical tagging, where 84 per cent of stars were tagged to an association of at least three stars. Over-plotted in Figure \ref{fig:chemtag} are the six most populated (\textgreater 19 members) chemically tagged groups with rms errors. It is clear from Figure~\ref{fig:chemtag} that OCs do not show any signs of a Na-O anti-correlation as a population. Together with the homogeneity discussion in Section \ref{inhomo}, we confirm the lack of Na-O anti-correlation in OCs.

\subsection{Open clusters and the field}

\begin{figure}
\centering
\includegraphics[width=9cm]{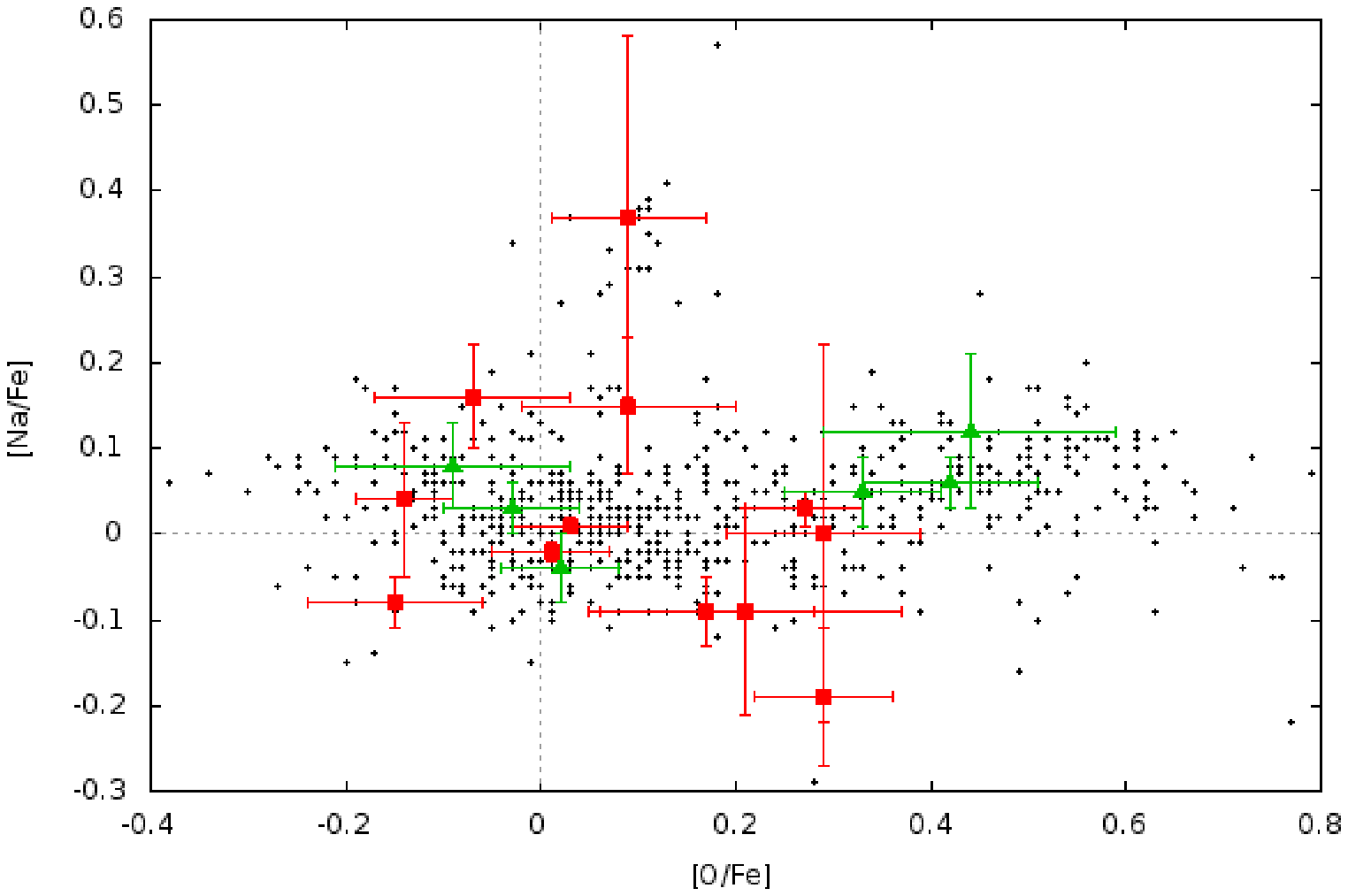}
\caption{Average normalized [Na/Fe] vs. [O/Fe] abundances with rms errors for open clusters (red squares) and the most populous chemically tagged groups from \protect\cite{chemtagdata} (green squares). Black dots are field stars from \protect\cite{bensby2014study}. Stars that showed inhomogeneity, as discussed in Section \protect\ref{inhomo}, were not taken into account for these averages, with the exception of NGC 6791.}
\label{fig:chemtag}
\end{figure}

In Figure~\ref{fig:chemtag}, the homogenised sample of open clusters match the disk very well. Clearly shown is a lack of any significant [Na/Fe] enhancement. Many open clusters in recent years have been noted to be significantly more Na-rich than the Galactic disk, hinting that present-day OCs were not a main contributor to the field. This was suggested notably by \cite{desilva2009oc} who did a similar study of OCs; however only LTE [Na/Fe] abundances were used (for the sake of homogeneity). In the current study, all [Na/Fe] abundances were adjusted to account for NLTE effects. This had the effect of lowering the OC [Na/Fe] abundances by 0.1--0.15 dex, bringing the open cluster sample in line with the Galactic disk sample.

\cite{chemtagdata} suggest that the chemically tagged groups in Figure~\ref{fig:chemtag} represent long dispersed coeval stellar clusters, with ages estimated to be between 1--10 Gyr, comparable with the OC ages. If true, this indicates that the homogenized sample of present-day bound open clusters matches this sample of entirely disrupted clusters, while being dynamically different in origin. Such clusters would have experienced high infant mortality rates  \citep[for which homogeneity is expected as per][]{krumholz2014}, while the OCs we observe now are bound and their relative contribution to the field is unclear from an observational point of view. Figure~\ref{fig:chemtag} shows certain overlap between the two star formation modes, suggesting that the chemical evolution was similar. In this case OCs adequately represent the dissipated clusters that formed the Galactic field. 

\subsection{Ba and Eu}

\begin{figure}
\centering
\includegraphics[width=9cm]{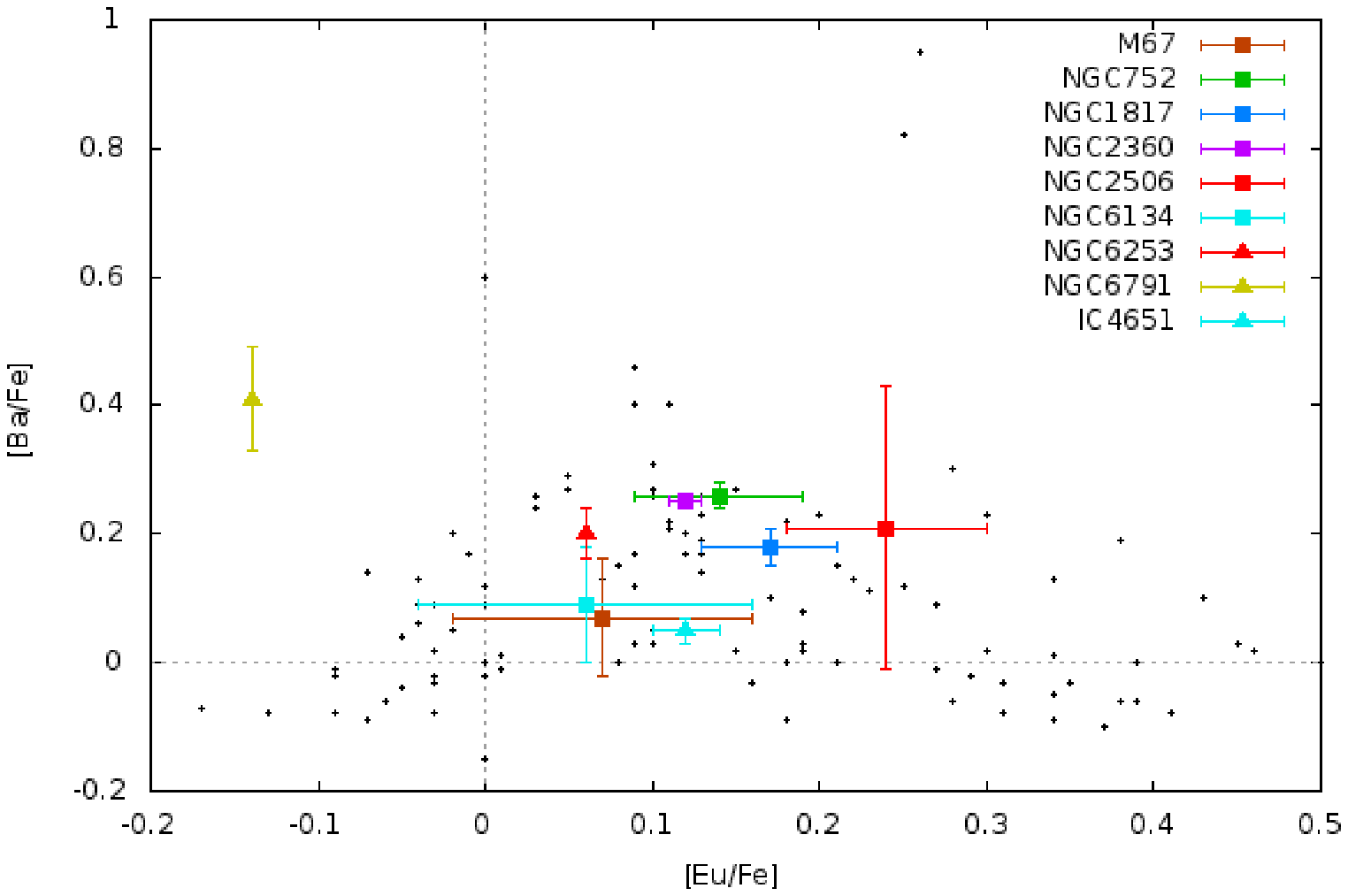}
\caption{Average normalized [Ba/Fe] vs. [Eu/Fe] abundances with rms errors for open clusters (squares and triangles). Black dots are field stars from \protect\cite{bensby2014study}.}
\label{fig:baeu}
\end{figure}

Figure~\ref{fig:baeu} shows the average [Ba/Fe] and [Eu/Fe] abundances of normalized open clusters compared with the local field sample of \cite{bensby2004study}, where all open clusters presented were found to be homogeneous in both Ba and Eu. With the exception of NGC 6791 which clearly occupies a separate position in the abundance space (and as stated in Section 3.1 it has many peculiar characteristics which may be the cause of its unusual Ba to Eu abundance ratio) all open clusters in the sample match the field very well, with no apparent trend in [Ba/Fe] vs. [Eu/Fe].

\cite{jacobson2013} observed a slight positive correlation between [Ba/Fe] and [Eu/Fe], suggesting improvement in s-process efficiency with r-process enrichment. Due to the large error in the Ba abundance of NGC 2506, this trend can be neither ruled out nor confirmed.

\section{Conclusions}

Open cluster abundance studies of Na, O, Ba and Eu were compiled from the literature in order to conduct a wide search for abundance patterns, particularly for the Na-O anti-correlation seen universally in GCs. In order to minimise the compounding of systematic errors, the data was homogenized as much as reasonable by adjusting a single solar reference and accounting for NLTE effects, variations in analysis, and choice of spectra lines.

The individual stars within each OC were found to be homogeneous in all presented chemical species. Any larger scatter (mainly in O) was found to be caused by large 
systematic errors in abundance analysis, deviations from LTE, and stellar parameters providing difficult observations of absorption lines. When compared to the [Na/Fe] and [O/Fe] abundances of NGC 2808, a template GC illustrating the Na-O anti-correlation, no similar trend was found in (mean) cluster values for the the OC sample. Thus we see no evidence for the GC O-Na correlation, either internally for stars in the OCs or globally, 
among the OC sample.

Once much of the systematics were addressed, the open cluster sample matched the abundances from the local field in all examined elements. 

The commonly noted Na-enhancement of OCs was not observed when NLTE corrections were applied, suggesting that this enhancement was an artefact of Na abundance analysis methods. The agreement between OCs and the field was particularly pronounced when the average OC abundances were compared with chemically tagged coeval groups from the field sample, which are thought to represent long disrupted star clusters having since formed part of the Galactic disk. This suggests that present-day OCs are representative of the star forming events that formed much of the Galactic disk, validating the traditional perspective that OCs are excellent probes into the chemical evolution of the Galactic disk. Large scale chemical tagging experiments such as the spectroscopic surveys of GALAH \citep{desilvainprep}, APOGEE \citep{apogee} and Gaia-ESO \citep[][which could particularly address many of our uncertainties with its systematic analysis of OCs]{gaia2} will help to provide a clearer picture of both the dynamic and chemical evolutionary history of the Galaxy.


\section*{Acknowledgements}

We acknowledge Ricardo Carrera for providing help with data collection and Karin Lind for providing the NLTE correction code. This research was supported under Australian Research Council’s Discovery Projects funding scheme (project numbers DP1095368 and DP120101815).


\bibliography{References}





\label{lastpage}

\end{document}